\documentclass[a4paper]{article}

\usepackage{INTERSPEECH2021}

\usepackage{booktabs}
\usepackage{multirow}
\usepackage{amsfonts}

\title{Speech2Video: Cross-Modal Distillation for Speech to Video Generation}

\name{Shijing Si, Jianzong Wang
, Xiaoyang Qu, Ning Cheng, Wenqi Wei, Xinghua Zhu and Jing Xiao}
\address{
    Ping An Technology (Shenzhen) Co., Ltd., China}
\email{jzwang@188.com}

\begin{document}

\maketitle

\begin{abstract}
This paper investigates a novel task of talking face video generation solely from speeches. 
The speech-to-video generation technique can spark interesting applications in entertainment, customer service, and human-computer-interaction industries.
Indeed, the timbre, accent and speed in speeches could contain rich information relevant to speakers' appearance.
The challenge mainly lies in disentangling the distinct visual attributes from audio signals.
In this article, we propose a light-weight, cross-modal distillation method to extract disentangled emotional and identity information from unlabelled video inputs.
The extracted features are then integrated by a generative adversarial network into talking face video clips.
With carefully crafted discriminators, the proposed framework achieves realistic generation results. 
Experiments with observed individuals demonstrated that the proposed framework captures the emotional expressions solely from speeches, and produces spontaneous facial motion in the video output.
Compared to the baseline method where speeches are combined with a static image of the speaker, the results of the proposed framework is almost indistinguishable.
User studies also show that the proposed method outperforms the existing algorithms in terms of emotion expression in the generated videos.

\end{abstract}
\section{Introduction}
\label{sec:intro}

\begin{figure}[htb]
  \centering
  \includegraphics[scale=0.4]{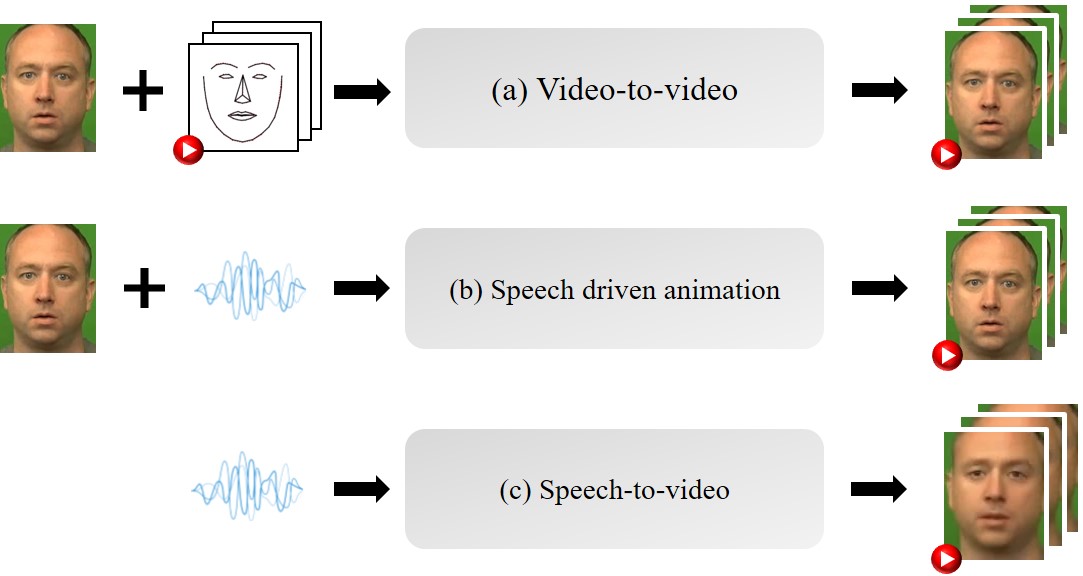}
  \caption{Comparison of different tasks of talking face video synthesis. a) Video-to-video synthesis with face landmark video and reference image as inputs \cite{zhou2019talking,mittal2019animating}. b) Speech driven animation leveraging given speeches and reference image \cite{vougioukas2019end,2019sdfa}. c) The proposed speech-to-video generation solely from speeches.}
  \label{fig:task_comp}
\end{figure}

Speeches are audio signals carrying affluent information of various domains \cite{richard2021audio,masood2021deepfakes}.
From a piece of speech, words and sentences are the linguistic contents it conveys. 
Emotions can be detected from the tone of a speech \cite{nagrani2020disentangled,wang2021generating}. 
Also, the speakers' identities are engraved in their voices, their accents and ways they talk \cite{cai2020speaker,yoon2020speech,lopez2021analysis}. 
Human brain can extract information on different aspects from a single piece of speech and recall or form imaginations of the identity of the potential speaker and also the speaker's facial movements when talking \cite{emre2020speech,huang2020fine,koumparoulis2020audio,prajwal2020lip,sinha2020identity,zeng2020talking}. 
In this paper, we propose an innovative task of realistic talking face generation from a single audio input. 

Prior to this paper, numerous works have been proposed to integrate visual and audio inputs to compose talking face videos of target identity \cite{wang2020mead,eskimez2020end,yang2020one,wang2020speech}.
Different settings of input sources is depicted in Fig.~\ref{fig:task_comp}.
In contrast to prior techniques, the proposed speech-to-video generator takes a segment of speech as the sole input.
By ``recollecting'' a face from learned audio features, the proposed framework produces a video with a face observed in the training process.
The results of speech2video is similar to that of speech driven animation.

The proposed task is challenging as well as intriguing.
Information about the speaker's identity and his/her spontaneous emotion is entangled in the speech signal, together with the linguistic contents.
The system must disentangle the identity and emotional attributes from the audio signal, and transfer these audio features into visual ones.
In this paper, we design a two-stage framework, speech2video, to its solution.
First, a cross-modal distillation module extracts identity and emotional features from unlabelled talking face videos.
Secondly, a generative adversarial network (GAN) \cite{goodfellow2014gan,creswell2018generative,karras2019style} with visual identity guidance is trained to compose the talking face video from the audio feature vectors.
Experimental results verifies the viability of the proposed task and solution.
Our generated videos of observed persons are realistic and emotionally accurate, similar to those of speech driven animation, even without any visual cues as input.
Results of unobserved persons do not have the same facial appearance as the ground truth, but is roughly consistent with respect to gender and age.

The contributions of this paper are:
\begin{itemize}
\item Propose a multi-modal distillation network to disentangle identity and emotional features from speech signals.
\item Devise a GAN structure with carefully crafted discriminators to generate talking face videos from speeches.
\item Demonstrate the viability of the proposed task, and evaluate the performance of the proposed video generator framework with extensive experiments.
\end{itemize}

\section{Methods}

\begin{figure*}[htbp]
  \centering
  \begin{minipage}{0.4\linewidth}
  \includegraphics[width=\textwidth]{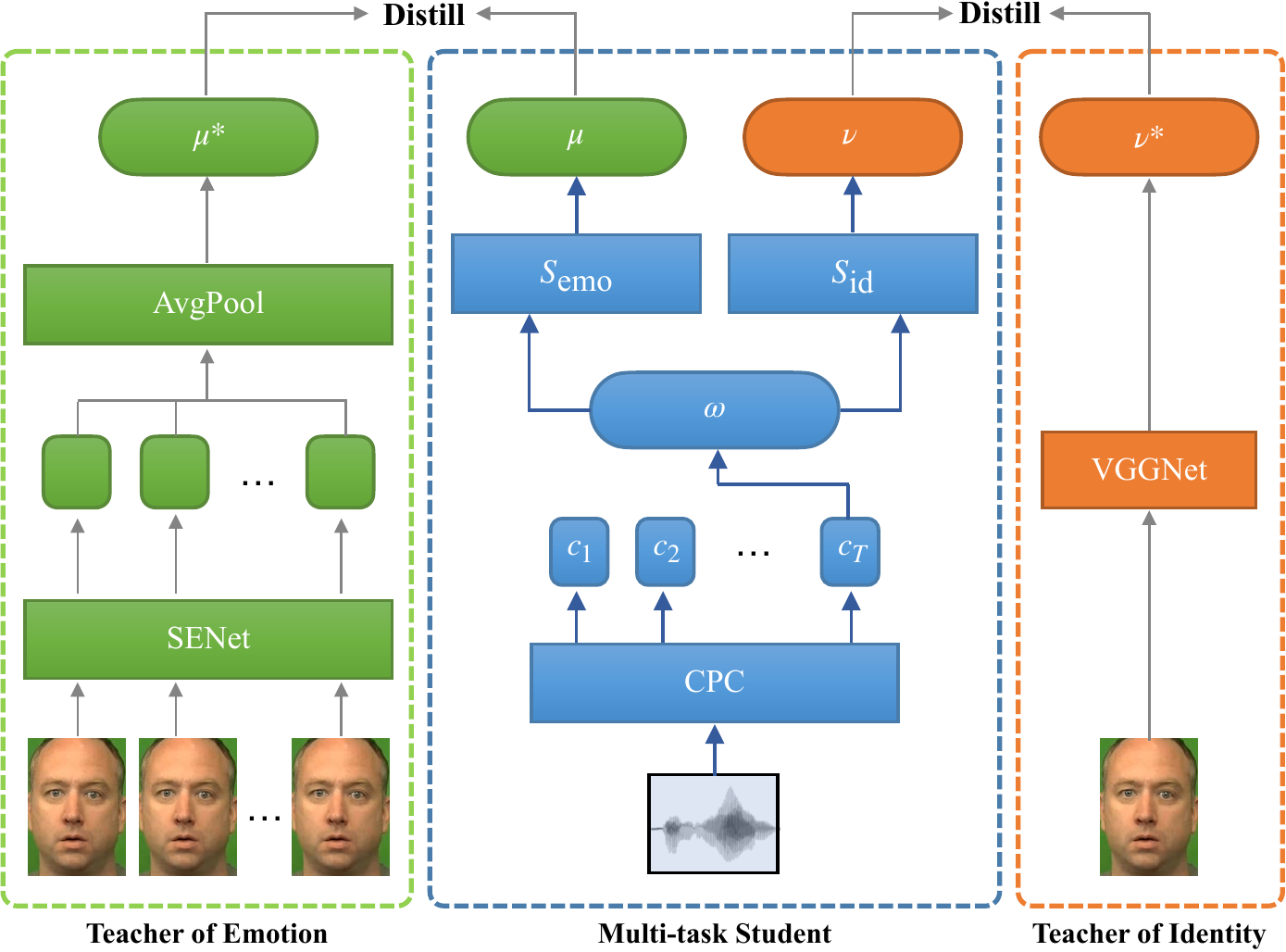}
  \end{minipage} \hspace{30pt}
  \begin{minipage}{0.45\linewidth}
  \includegraphics[width=\textwidth]{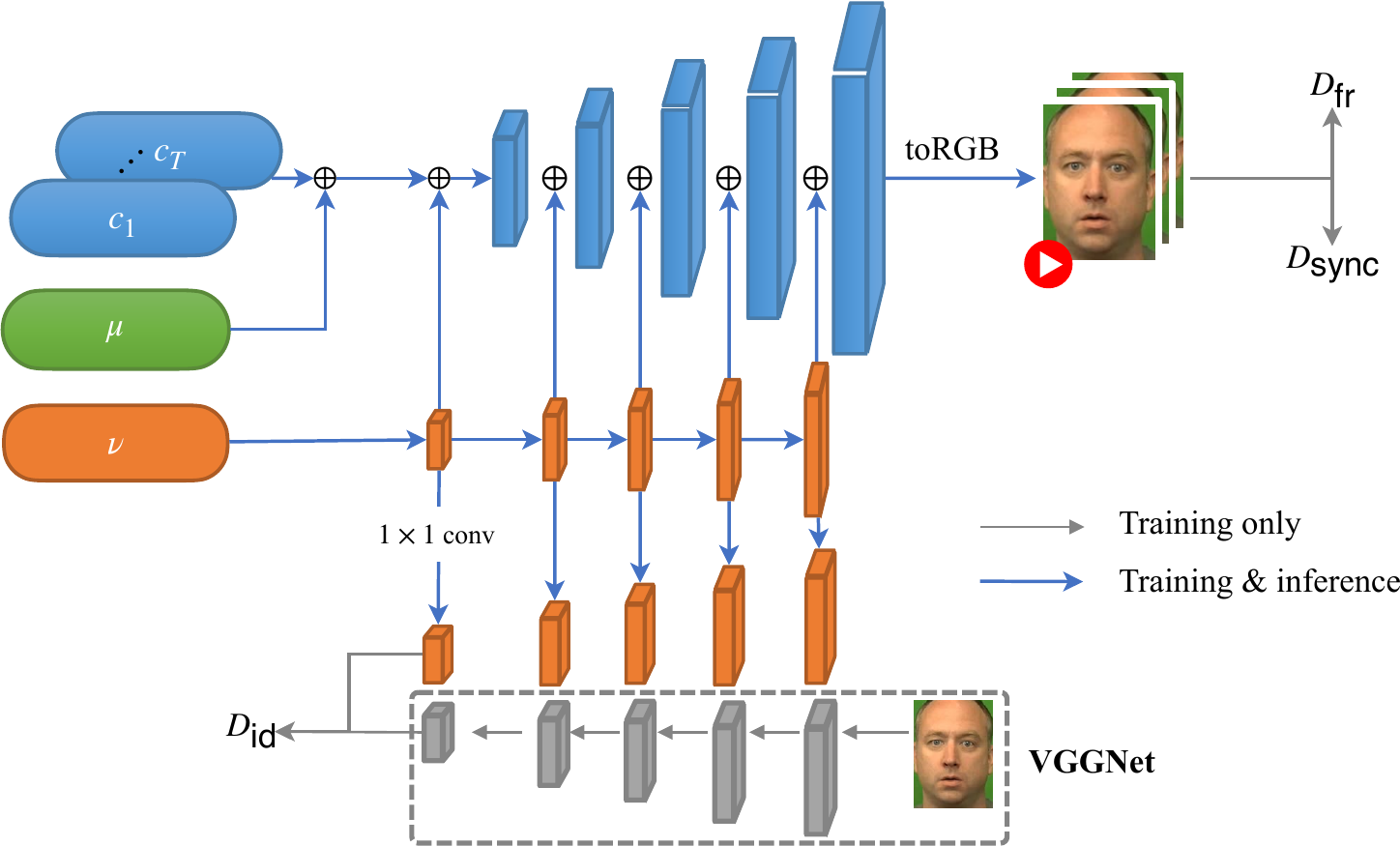}
  \end{minipage}
  \caption{The speech2video framework. Left: Proposed feature disentangling model. Right: Adversarial frame generator with disentangled intermediate features. The $\oplus$ symbol represents feature concatenation and transposed-convolution, followed by two $1\times 1$ convolution layers.}
  \label{fig:framework}
\end{figure*}

The overall framework of speech2video is illustrated in Fig. \ref{fig:framework}.
The comprehensive procedure composes of two stages.
In the first stage, the personal identity and emotional features are disentangled from the speech representation.
The representation disentangling module is trained from unlabelled talking face videos, using distillation learning techniques.
In the second stage, an adversarial composer gathers the temporal speech representation as well as the identity and emotional features, and generates the talking face frames.
The representation disentangler and adversarial composer are trained separately.
Detailed description of their structures is elaborated in the following subsections.

\subsection{Speech Representation Disentangling}

\newcommand{\cpcvec}{\mathbf{\omega}}
\newcommand{\emovec}{\mathbf{\mu}}
\newcommand{\idvec}{\mathbf{\nu}}
\newcommand{\audioin}{\mathbf{\alpha}}
\newcommand{\videoin}{\mathbf{\phi}}

Speeches carry affluent information about speakers' identity.
Setting the linguistic contents aside, human receivers have the natural ability to recognize the speakers' identity and emotion from the speeches alone.
However, for a neural network, these attributes are too entangled to understand.
In order to delineate and preserve the identity and emotional characteristics in a speech, we must first disentangle these features from the original speech.

The disentangling task is non-trivial.
For one thing, the input signal is in the audio domain, while target features are visual, requiring cross-domain information encoding.
For another, the disentangling module must extract multiple features, namely the identity feature and emotional feature, at the same time.
Furthermore, in order to maximize the applicable source of training samples, the disentangling module is optimized in an unsupervised manner.

In this section, a cross-modal distillation network is proposed as the disentangling module (Fig. \ref{fig:framework}a).
At the training stage, an unlabelled talking face video $V = \{(\alpha_t, \phi_t), t=1, ..., T\}$ is provided as input.
The targeted representation extractor learns only from the audio signals $\audioin = \{\alpha_1, ..., \alpha_T\}$.
The visual segments $\videoin = \{\phi_1, ..., \phi_T\}$ in the video provide supervisory information to the audio representation through a distillation network.
By assigning different teacher networks, the identity and emotional features, $\idvec$ and $\emovec$, are delineated through the respective distillation process simultaneously.

Specifically, a comprehensive representation of audio features is extracted by a pre-trained contrastive prediction coding (CPC) module \cite{oord2018representation} from the speech.
In CPC, the audio sequence $\alpha$ is first recurrently encoded by a nonlinear encoder $g_\text{enc}$ into latent embedding $z_t, t=1,...,T$.
Then an auto-regressive model $g_\text{ar}$ summarizes all latent embedding up to time $t$ and produces context latent representation
$c_t = g_\text{ar}(z_{\leq t})$.
CPC has been shown to excel in downstream tasks like speaker classification and phone classification.
The temporal vector $C = \{c_1, ..., c_T\}$ is used to generate frame sequences, as elaborated in the next section.
Additionally, we take $\cpcvec = c_T$
as a dimensionality-reduced interpretation of the intact information in the audio signal.

Two student networks, $S_\text{emo}$ and $S_\text{id}$, read from $\cpcvec$ and distill emotional and identity features, respectively.
Let $\idvec := S_\text{id}(\cpcvec)$ and $\emovec := S_\text{emo}(\cpcvec)$ denote the distilled feature vectors.
In accordance with the student networks, two teacher networks, $T_\text{emo}$ and $T_\text{id}$, are defined to supervise the student networks on knowledge extraction .
The $T_\text{id}$ module adopts a VGGNet structure pre-trained on the VGGFace dataset for face recognition.
Through the VGGNet, a 4,096-dimensional identity feature vector is extracted from the first frame of the video, $\idvec^* = T_\text{id} \left( \videoin_0 \right)$.
On the other hand, as emotional motion changes from frame to frame, it is insufficient to observe a single frame. 
The $T_\text{emo}$ module takes $K$ randomly sampled frames $\pi_1, ..., \pi_k$ from $\videoin$, and extracts the corresponding emotion feature vectors through a Squeeze and Excitation Network (SENet).
The target feature for $\emovec$ is the average pooling of these samples, i.e.,
\begin{equation}
    \emovec^* = \frac{1}{K} \sum_{k=1}^K T_\text{emo} \left(\videoin_{\pi_k}\right).
\end{equation}
The choice of the two teacher networks can be tuned to dataset characteristics and computation power.
In this paper, we adopt SENet and VGGNet for simplicity and extensive proofs of their performance in the respective fields.
The teacher networks are pre-trained on open-source facial image dataset VGGFace \cite{vgg_face,haque2019grayscale} and FERplus \cite{barsoum2016training}.
Parameters of $S_\text{id}$ and $S_\text{emo}$ are optimized simultaneously, to minimize the joint multi-task distillation loss \cite{Parkhi15},
\begin{align}
\begin{split}
    L_1 = & \lambda\| \bar{\emovec} - \bar{\emovec}^* \|^2 + l_\text{distill}\left(\emovec, \emovec^* \right) + \\
    & \lambda\| \bar{\idvec} - \bar{\idvec}^* \|^2 + l_\text{distill}\left(\idvec, \idvec^* \right),
\end{split}
\label{eq:loss1}
\end{align}
where $\lambda=0.025$ and $\bar{(\cdot)}$ denotes the normalized vectors. 
$l_\text{distill}$ stands for the distillation loss, i.e., the softmax cross-entropy
\begin{equation}
    l_\text{distill}(\mathbf{x}, \mathbf{y}) = -\sum_i \text{softmax}(\mathbf{x})_i \log (\text{softmax}(\mathbf{y})_i).
\end{equation}

\subsection{Adversarial Video Composition}

To alleviate the complexity of speech-to-video generation, the proposed framework is divided into two separate stages.
The cross-modal encoding of speech signal has been covered in the last section.
In this section, an adversarial decoder is proposed to transform the encoded vectors into a talking face video.

From the encoder, we have obtained 3 feature vectors, namely the frame feature sequence $C$, the emotion feature $\emovec$ and the identity feature $\idvec$.
For every frame $c_t \in C$, the emotion feature is directly concatenated for transposed convolution.
On the other hand, the identity feature $\idvec$ is separately handled for enhanced feature supervision.

Specifically, $\idvec$ goes through 3 transposed convolution layers, each doubling its width and height.
For every intermediate feature map, a $1\times 1$ convolution aligns its shape to the corresponding layer in a VGGNet.
The pretrained VGGNet features of a single facial image is adopted to supervise the generation of high-resolution identity feature maps by an adversarial loss, to be detailed in the remainder of this section.
The inflated identity features are concatenated to the corresponding layers of the frame generator, and input to the following transposed convolutions.
We have found that processing the identity feature separately greatly enhanced the individual distinction in the final video frames.
The frame decoder $F$ is trained to optimize the combination of a number of target functions, as elaborated below.

\textbf{Adversarial losses.} 
The frame decoder $F$ is a generator that maps audio features to video frames.
Generative Adversarial Networks (GANs) supplement the generator with a competing discriminator, to train the generator models unsupervisedly.
In the proposed framework, 3 discriminators are devised to audit the decoder $F$ from different perspectives, namely, identity preservation, frame authenticity, and synchronization.
These discriminators are denoted as $D_\text{ID}$, $D_\text{fr}$, $D_\text{sync}$, respectively.
The adversarial losses used in our experiments are Least Squares GAN (LSGAN) losses.

$D_\text{ID}$ discriminates the inflated identity feature maps $\tilde{\idvec} = \{\tilde{\idvec}^{(i)}, i=1,2,3\}$ against the VGG features $\mathbf{V}=\{\text{VGG}^{(i)}(I), i=1,2,3\}$ of a static facial image $I$ of the speaker, i.e.,
\begin{equation}
l_\text{adv}^\text{ID} = \frac{1}{2} \mathbb{E} \left[ (D_\text{ID} (\tilde{\idvec}) -1 \right]^2  +
\frac{1}{2} \mathbb{E} \left[ D_\text{ID} (\mathbf{V}) \right]^2.
\end{equation}

$D_\text{fr}$ discriminates between the generated and real video frames.
The corresponding adversarial loss is 
\begin{equation}
l_\text{adv}^\text{fr} = \frac{1}{2}\sum_{t=1}^T \Big(\mathbb{E} \left[ (D_\text{fr} (F(c_t, \idvec, \emovec) -1 \right]^2  +
\mathbb{E} \left[ D_\text{fr} (\videoin_t) \right]^2\Big).
\end{equation}

Last but not least, $D_\text{sync}$ ensures the synchronization between audio and video frames.
For this discriminator, time steps $\tau$ and $\tau'$ are sampled from $\{1, ..., T\}$, $\tau \neq \tau'$.
$D_\text{sync}$ takes a pair of audio feature and video frame as input.
The synchronous adversarial loss encourages synchronized audio-video pairs while punishing the asynchronous ones, i.e.,
\begin{equation}
\begin{split}
l_\text{adv}^\text{sync} = &\mathbb{E} \left[ (D_\text{sync} (c_\tau, \videoin_\tau) \right]^2  + \\
&\frac{1}{2} \mathbb{E} \left[ (D_\text{sync} (c_{\tau'}, \videoin_\tau) -1 \right]^2  + \\
&\frac{1}{2} \mathbb{E} \left[ D_\text{sync} (c_\tau, F(c_\tau, \idvec, \emovec)) -1 \right]^2.
\end{split}
\end{equation}

The overall adversarial loss is given by
\begin{equation}
    L_2^\text{adv} = l_\text{adv}^\text{ID} + l_\text{adv}^\text{fr} + l_\text{adv}^\text{sync}.
\end{equation}

\textbf{Frame similarity loss.}
In addition to the adversarial losses, a pixel-wise similarity loss is imposed on the generated frames.
Particularly, the facial appearance of a talking person remains largely identical from the nose up, that occupies roughly the upper half of the frame image.
Therefore, the frame similarity loss is defined as
\begin{equation}
    L_2^\text{sim} = \sum_{p\in [0, W]\times[H/2, H]} | F^p (c_t, \idvec, \emovec) - \videoin_t^p |,
\end{equation}
where $W$ and $H$ are the width and height of the video frames, respectively.
$F^p$ and ${\videoin}_t^p$ stands for the pixel value of the corresponding frame image at position $p$.

\textbf{Gradient loss.}
As suggested by \cite{kou2015gradient}, a gradient loss is also applied to alleviate the blurriness caused by L1 frame similarity function.
\begin{equation}
    L_2^\text{grad} = \left|\nabla \psi(F(c_t, \idvec, \emovec) - \nabla \psi(\videoin_t)) \right|,
\end{equation}
where $\psi(\cdot)$ is a smoothing filter, $\nabla$ is the gradient filter.

In summary, the total loss for the frame decoder $F$ is
\begin{equation}
    L_2 = L_2^\text{adv} + L_2^\text{sim} + L_2^\text{grad}.
\end{equation}


\section{Experiments}

\subsection{Implementation Details}

The proposed speech2video framework is implemented on PyTorch with a single NVIDIA Tesla V100 GPU.
The learning rate of the generator, identity discriminator, frame discriminator and naturalness discriminator is set to $3\times 10^{-4}$, $3\times 10^{-4}$, $1\times 10^{-4}$ and $1\times 10^{-5}$, respectively. 
RMSProp optimizer is adopted for all the training. 
The disentanglement and frame decoder modules are trained for 100 epochs respectively.


The proposed methods are experimented with two open datasets of talking face videos, Crowd Sourced Emotional Multimodal Actors Dataset (CREMA-D) \cite{cao2014crema} and VoxCeleb2 \cite{voxceleb2}.
\textit{CREMA-D} contains 7,442 clips uttered by 91 ethnically-diverse actors(48 male, 43 female). 
Each speaker utters 12 sentences in 6 different emotions (Anger, Disgust, Fear, Happy, Neutral, Sad). 
This dataset is challenging because facial movements of the speaker under certain emotions are expressive or even exaggerated. 
The audio sample rate and video sample rate of this dataset is 16kHz and 30fps respectively. 
The dataset is divided by the proportion $70\%$,$15\%$, and $15\%$ for training, validation and testing, respectively.
\textit{VoxCeleb2} contains more than 1 million utterances of  6,112 celebrities extracted from YouTube videos. 
The training and testing sets are given by the authors of the dataset.

In the following experiments, performance of Speech Driven Facial Animation (SDFA) \cite {vougioukas2019end,2019sdfa} is compared with the proposed framework as a baseline.
SDFA utilizes a static facial image and a piece of raw waveform speech to synthesize a talking face video and has achieved realistic reconstruction of facial features in the generated results.
Furthermore, for a baseline of facial video generation with only a voice input, we combine the voice encoder of speech2face \cite{oh2019speech2face} with the video generator of SDFA, to form the S2F+SDFA model.


\subsection{Qualitative Results}

\begin{figure}[htb]
  \centering
  \includegraphics[width=0.99\linewidth]{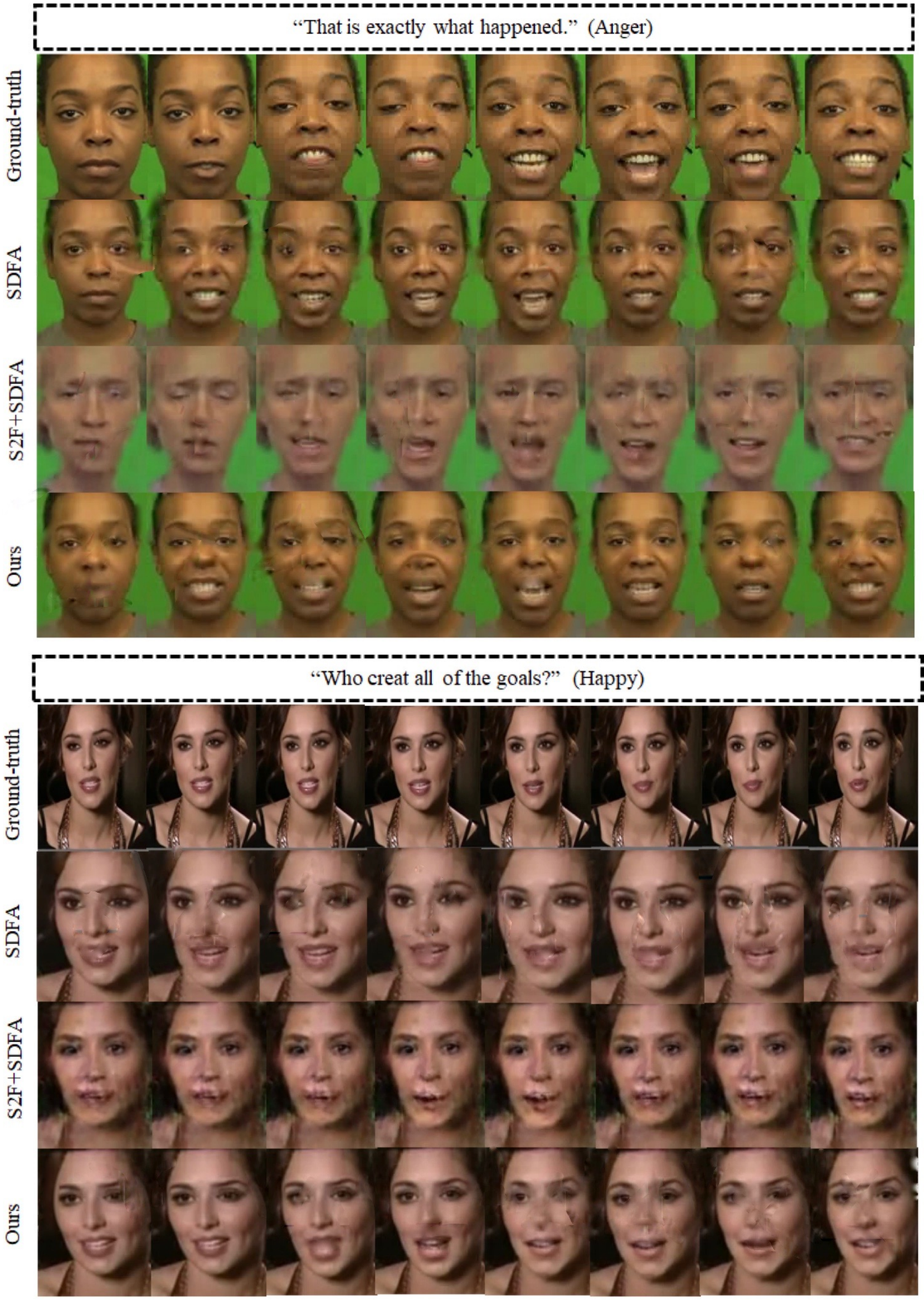}
  \caption{Qualitative result comparison of our system with two baseline models and the ground-truth on two input speeches. For each input utterance, Ground-truth is in the first row, followed by SDFA and S2F+SDFA, with our method in the last row. Our result not only possesses accurate general features but also abundant facial details.}
  \label{fig:comparison}
\end{figure}

Figure \ref{fig:comparison} demonstrates the performance of designed model comparing to the baseline methods, SDFA and S2F+SDFA. 
It can be seen that the S2F+SDFA network yields the worst performance with large identity error and blurry frames. 
This is because unlike the designed model with adversarial training in the identity reconstruction process, modified speech2face model only leverages a normalization loss function on high dimensional embedding to supervise the identity which is insufficient. 
Furthermore, the SDFA model is designed to have a reference image as input. 
However, in S2F+SDFA network, the reference image is substituted by high dimensional facial features. 
This lacks of detailed information for the decoding process which results in bad quality frames. 
On the other hand, the proposed model obtains competitive results comparing to the state-of-art SDFA model, though leveraging no reference images. Identity and facial details of the speaker are accurately reconstructed. 
Though, the SDFA model contains more texture details, such as light and shade, which are carried in the reference image. 
In summary, the experiments well demonstrated the competitive performance and the feasibility of the proposed methodology.

\begin{figure}[htb]
  \centering
  \includegraphics[width=0.9\linewidth]{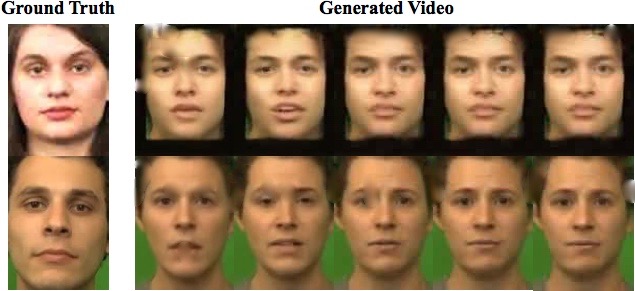}
  \caption{Imaginary facial video generation by our method from unobserved voice inputs.}
  \label{fig:imagine}
\end{figure}

In the above experiments, the proposed framework is tasked to ``recollect'' faces from observed voices.
It would be also interesting to see what we can achieve if an unobserved voice is given to the speech2video framework.
Fig. \ref{fig:imagine} illustrates two cases of un-observed voice inputs.
As it shows, the appearance of the generated faces are different from the ground truth.
But the speakers' gender, complexion and age are roughly preserved in the results.

\subsection{Quantitative Results}

Quantitatively, we evaluate proposed model based on the following aspects: correctness of identity reconstruction, the quality of generated videos and the audio-visual synchronization. 
The corresponding metrics are structural similarity (SSIM), peak signal-to-noise ratio (PSNR) and AV confidence \cite{chung2016out} indices, respectively.

The quantitative results of the proposed model compared with baseline methods on the aforementioned metrics are shown in table \ref{tab:quantative_comp}. 
The proposed model scores below the SDFA model but higher than the S2F+SDFA network which follows the expectation. 
SSIM and PSNR metrics represent the distance between the generated and ground-truth frame in various aspects. 
The proposed system does not use ground-truth image as reference.
Because of the resulting diversity of GAN based approach, it is impossible to reconstruct facial features as perfectly without any additional image as references. 

\begin{table}[t]
\centering
\small
\renewcommand\tabcolsep{2.4pt}
\caption{Quantitative comparison of the proposed method with baseline methods.}
\begin{tabular}{|c|c|c|c|c|}
\hline
\textbf{Dataset}           & \textbf{Method} & \textbf{SSIM} & \textbf{PSNR} & \textbf{Confidence} \\ \hline
\multirow{4}{*}{CREMA-D}   & SDFA            & 0.705         & 23.568        & 5.4                    \\ \cline{2-5} 
                           & S2F+SDFA        & 0.519         & 20.104        & 4.9                    \\ \cline{2-5} 
                           & SDFA(mismatch)  & 0.521         & 20.024        & 5.0                    \\ \cline{2-5} 
                           & speech2video        & 0.541         & 20.540        & 5.1                    \\ \hline
\multirow{4}{*}{VOXCELEB2} & SDFA            & 0.7342        & 25.121        & 5.8                    \\ \cline{2-5} 
                           & S2F+SDFA        & 0.5643        & 20.413        & 4.9                    \\ \cline{2-5} 
                           & SDFA(mismatch)  & 0.5612        & 20.126        & 5.0                    \\ \cline{2-5} 
                           & speech2video        & 0.5707        & 20.891        & 5.1                    \\ \hline
\end{tabular}
\label{tab:quantative_comp}
\end{table}

\section{Conclusion and Future Work}
This paper explores a novel model to synthesize talking face video solely from a single audio. 
With the help of cross-modal distillation, the model extracts embedding vectors representing emotions and speaker identities, and generates face videos through an adversarial framework. 
The design of multiple discriminators ensures the naturalness and fluency of the generated video. Through a series of experiments, our designed model shows persuasive results. 

\section{Acknowledgements}
This work is supported by National Key Research and Development Program of China under grant No.2018YFB0204403, No.2017YFB1401202 and No.2018YFB1003500. Corresponding author is Jianzong Wang from Ping An Technology (Shenzhen) Co., Ltd.

\newpage
\bibliographystyle{IEEEtran}
\bibliography{ijcai20}

\end{document}